\documentclass[10pt,fleqn,twocolumn,aps,pre,groupedaddress,a4paper,floatfix]{revtex4}

\usepackage{amsmath}
\usepackage{graphicx,psfrag}
\usepackage{multirow}
\usepackage{graphics}
\usepackage{amsfonts}
\usepackage{amssymb}
\usepackage{subfigure}

\begin{document}
 \title{Information and entropy in quantum Brownian motion:\\
  Thermodynamic entropy versus von Neumann entropy}
 \author{Christian H\"orhammer}
 \email{christian.hoerhammer@uni-bayreuth.de}
 \affiliation{Theoretische Physik I, Universit\"at Bayreuth, D-95440 Bayreuth, Germany}
 \author{Helmut B\"uttner}
 \affiliation{Theoretische Physik I, Universit\"at Bayreuth, D-95440 Bayreuth, Germany}
 \date{\today}
 \begin{abstract}
We compare the thermodynamic entropy of a quantum Brownian
oscillator derived from the partition function of the subsystem with
the von Neumann entropy of its reduced density matrix. At low
temperatures we find deviations between these two entropies which
are due to the fact that the Brownian particle and its environment
are entangled. We give an explanation for these findings and point
out that these deviations become important in cases where statements
about the information capacity of the subsystem are associated with
thermodynamic properties, as it is the case for the Landauer principle.
\end{abstract}
\maketitle
\section{Introduction}
Quantum systems that are not perfectly isolated from their
environment are subject to dissipation and decoherence. Within the
theory of quantum dissipative systems \cite{weiss, breuer, dittrich}
models with system-reservoir interaction such as Quantum Brownian
motion \cite{caldeira1, ford1} have been developed to describe these
phenomena. The results are of great interest for applications in the
field of mesoscopic systems, quantum computation \cite{nielsen} and
quantum information theory of continuous variable systems \cite{braunstein, cerf}. \\

In last years the behavior of open quantum systems has also raised questions about the validity of
fundamental laws of thermodynamics \cite{sheehan, nieuw, NA1} and has again been addressed by
recently published papers \cite{ford0a, ford0b, haenggi2}. The central point in this discussion is the proper
a definition of the thermodynamic
quantities of interest, especially the entropy. For this
reason we want to provide a comprehensive overview of the two most
common measures of the entropy of an open quantum system -- the von
Neumann entropy and the thermodynamic entropy -- and point out that these two quantities might
deviate in the quantum Brownian motion model.\\

The von Neumann entropy of the subsystem can be derived from the
reduced density matrix \cite{ohya}. In quantum information theory it
is a commonly used measure of entanglement and decoherence
\cite{braunstein} and can be understood as a measure of the amount
of information gained by a unique quantum measurement of
localization. The thermodynamic entropy, on the other hand, is
associated with the amount of heat contained in the subsystem. Heat
is defined as the part of the internal energy of a system which
cannot be transferred into mechanical work. The thermodynamic
entropy can be calculated from the
partition function of the system and its free energy respectively.\\

As long as the application of the von Neumann entropy is restricted to
an information-theoretical context and the thermodynamic entropy is
applied when examining the validity of the thermodynamic laws no
interpretational problems will arise. The situation is different in
cases where statements about the information content of a system are
connected to thermodynamic properties, as
will be shown in the following.\\

This paper is organized as follows. In the first part of this paper we give a short review of the
Caldeira-Leggett model of quantum Brownian motion which is the basis
for discussion. In the main part the thermodynamic entropy and the von
Neumann entropy are calculated and compared. Finally we discuss the
significance of the deviations with regard to an information-theoretical point of view, namely by means of the Landauer principle.

\section{Caldeira-Leggett model of quantum Brownian Motion}
The statistical properties of the stationary state are discussed here on the
basis of the Caldeira-Leggett model \cite{caldeira1,
caldeira2, caldeira3} of quantum Brownian motion,
often referred to as independent oscillator model \cite{ford4, ford3}. It is a system-plus-reservoir model where the total Hamiltonian consists of three parts
\begin{equation}
H=H_s+H_b+H_{\rm int},
\end{equation}
with $H_s$ as Hamiltonian of the subsystem which interacts via the Hamiltonian $H_{\rm int}$
with a bath that is described by a collection of a large number $N$ of harmonic oscillators $H_b=\sum_i\hbar\omega_i(b^{\dagger}b+1)$.
In detail the Hamiltonian of the Caldeira Leggett model is given by
{\mathindent0.5cm \begin{equation}
\label{Hcl}H=\frac{p^2}{2m}+V(q)+\sum_{i=1}^N\left[\frac{p_i^2}{2m_i}+\frac{m_i\omega_i^2}{2}
\left(x_i-\frac{c_iq}{m_i\omega_i^2}\right)^2\right],
\end{equation}}
where $q$ and $p$ are the Heisenberg-operators of coordinate and momentum of the Brownian oscillator moving in an harmonic potential $V(q)=\frac{1}{2}m\omega_0^2q^2$ and coupled to a bath of $N$ independent harmonic oscillators with variables $x_i$, $p_i$ and equidistant frequencies $\omega_i=i\Delta$. The bath is characterized by its spectral density
\begin{equation}\label{Jdrude}
J(\omega)=\pi\sum_{i=1}^N\frac{c_i^2}{2m\omega_i}\delta(\omega-\omega_i)=\frac{\gamma\omega\Gamma^2}{\omega^2+\Gamma^2}\end{equation}
with cut-off frequency $\Gamma=N\Delta$ and system-bath coupling strength $\gamma$.
The interaction is bilinear in the coordinates $q$ and $x_i$ of the subsystem and the bath respectively with coupling parameters $c_i=\sqrt{2\Delta m_i\omega_i J(\omega_i) /\pi}$. The self-interaction term
(proportional to $q^2$) in the Hamiltonian
\begin{equation}\label{Hint}
H_{\rm
int}=\sum_i\left[-c_ix_iq+\frac{c_i^2}{2m_i\omega_i^2}q^2\right]
\end{equation}
renormalizes the oscillator potential to ensure that the observable frequency is close to bare oscillator frequency $\omega_0$.
From the Hamiltonian \eqref{Hcl} the Heisenberg equations of motion
for the operators $q$ and $p$ and the bath variables $x_i$, $p_i$
are received. By eliminating the bath degrees of freedom the quantum
Langevin equation \cite{ford1, ford2} of a particle moving in the
potential $V(q)$ can be derived:
\begin{equation}\label{langevinBM} m\ddot q(t) + \frac{dV(q)}{dq} +
\int_0^t{dt'\gamma(t-t') \dot q(t')} = \eta(t) -
q(0)\gamma(t),\end{equation} with damping kernel
\begin{equation}\label{dkernel}
\gamma(t)=\frac{2}{\pi}\int_0^{\infty}d\omega\frac{J(\omega)}{\omega}\cos\omega
t=\gamma\Gamma e^{-\Gamma t}
\end{equation} and the noise term $\eta(t)$ characterized by its
correlation function \cite{gardiner}
\begin{equation}\label{nkernel}K(t-t')=\frac{\hbar}{\pi}
\int_0^\infty d\omega J(\omega)
\coth(\frac{1}{2}\beta\hbar\omega)\cos\omega(t-t').\end{equation}
The stationary state of the Brownian particle is fully characterized
the reduced density matrix $\rho_s$ of the subsystem \cite{weiss,
haenggi1}
\begin{equation}\label{roheq}\langle q|\rho_s|q'\rangle=\frac{1}{\sqrt{2\pi \langle q^2 \rangle}} \exp\left[-\frac{(q+q')^2}{8\langle q^2 \rangle}-\frac{(q-q')^2}{2
\hbar^2/\langle p^2 \rangle}\right],\end{equation} where the
equilibrium fluctuations $\langle q^2\rangle$, $\langle p^2\rangle$
and the systems response function $\tilde\chi(\omega)=\left[m\omega_0^2-m\omega-i\omega\tilde\gamma(\omega)\right]^{-1}$
with  $\tilde \gamma (\omega)=\int_0^{\infty} \gamma(t)e^{i\omega t}$ are
connected by the quantum fluctuation-dissipation-theorem which gives
\begin{eqnarray}\label{xxchi}\langle q^2\rangle&=&\frac{\hbar}{\pi}\int_{0}^{\infty}d\omega  \coth(\frac{1}{2} \beta \hbar \omega)
\mbox{Im}\{\tilde\chi(\omega)\},\\
\label{ppchi}\langle p^2\rangle&=&\frac{\hbar}{\pi}\int_{0}^{\infty}d\omega\, m^2\omega^2 \coth(\frac{1}{2} \beta \hbar \omega)
\mbox{Im}\{\tilde\chi(\omega)\}\end{eqnarray}
for the stationary correlations.

\section{Calculation of entropies}

\subsection{Partition function and thermodynamic entropy}
Starting point in the calculation of thermodynamic quantities is the
partition function. For dissipative quantum systems such as the QBM model it can be derived by influence functional path integral methods \cite{grabert},
\begin{equation}\label{Zsum}
\mathcal Z=\oint\mathcal{D}[q(t)]e^{-S_{\rm
eff}^E[q(t)]/\hbar}
\end{equation}
with effective Euclidian action $S_{\rm
eff}^E[q(t)]$ and is given by
\begin{equation}
\mathcal{Z}=\frac{1}{\beta\hbar\omega_0}\prod_{n=1}^{\infty}\frac{\nu_n^2}{\omega_0^2+\nu_n^2+\nu_n\hat\gamma(\nu_n)},
\end{equation}
with Laplace transform $\hat \gamma(z)$ of \eqref{dkernel} and
Matsubara frequencies $\nu_n=2\pi n/\hbar\beta$, $n=\pm 1,\pm2,...$.
For the Drude model \eqref{Jdrude} the partition function can be
rewritten in the form
\begin{equation}
\mathcal{Z}=\frac{\omega_0}{2\pi\nu}\frac{\Gamma_e(\lambda_1/\nu)\Gamma_e(\lambda_2/\nu)\Gamma_e(\lambda_3/\nu)}{\Gamma_e(\Gamma/\nu)},
\end{equation}
with the Gamma function $\Gamma_e(x)$, $\nu=|\nu_1|$ and $\lambda_j$, $j=1,..,3$ as three characteristic frequencies (poles) of the non-Markovian damped harmonic oscillator. The free
energy
\begin{equation}\label{Fthermo}
F=-kT\ln
\mathcal{Z}=kT\left[\ln\frac{\Gamma_e(\Gamma/\nu)}{\omega_0/(2\pi\nu)}-
\sum\limits_{j=1}^3\ln\Gamma_e(\lambda_j/\nu)\right]
\end{equation}
can also be written in the form \cite{ford3, ford0a}
\begin{equation}\label{Fp}
F(T)=\frac{1}{\pi}\int_0^{\infty}d\omega\,
f(\omega,T)\mbox{Im}\left\{\frac{d\ln\tilde
\chi(\omega)}{d\omega}\right\},
\end{equation}
with
$f(\omega,T)=kT\ln\left[2\sinh\left(\frac{1}{2}\beta\hbar\omega\right)\right]$
as free energy of the uncoupled harmonic oscillator. The
thermodynamic entropy
\begin{equation}\label{Sthermo} S(T)=-\frac{\partial F}{\partial
T}=\frac{1}{\pi}\int_0^{\infty}d\omega\,
s(\omega,T)\mbox{Im}\left\{\frac{d\ln \tilde
\chi(\omega)}{d\omega}\right\},\end{equation} with $s(\omega,
T)=\frac{\beta\hbar\omega}{e^{\beta\hbar\omega}-1}-\ln\left(1-e^{-\beta\hbar\omega}\right)$
as well as the internal energy $U=F+TS=-\frac{\partial}{\partial\beta}\ln\mathcal{Z}$ and the specific heat $C=T\frac{\partial
S}{\partial T}$ can then be calculated straight forward. At zero bath temperature the entropy vanishes $\lim_{T\to0}S(T)= 0$ as required by
the Third Law of thermodynamics \cite{ford0b, haenggi2}. In case of
an ohmic heat bath, the entropy vanishes linearly with $T$. A interpretation of the free energy \eqref{Fp} will be given in section V within the discussion on cyclic thermodynamic processes.

\subsection{Ground state energetics at zero temperature}
Despite the fact that the entropy \eqref{Sthermo} vanishes for $T\to
0$ the energy of the subsystem fluctuates. Considering the
subsystems Hamiltonian $H_s$ as an observable of interest,
projective measurements of $H_s$ can find the system in higher
energy states even at zero temperature \cite{buettiker1,buettiker2}.
\begin{figure}[t]
\begin{center}
\psfrag{n}[c]{n}\psfrag{m}[c]{m} \psfrag{r}[c]{$|\rho_{\rm nm}|$}
\psfrag{2}[c]{\rm 1}\psfrag{4}[c]{3}\psfrag{6}[c]{\rm 5}
 \includegraphics[width=6cm,clip]{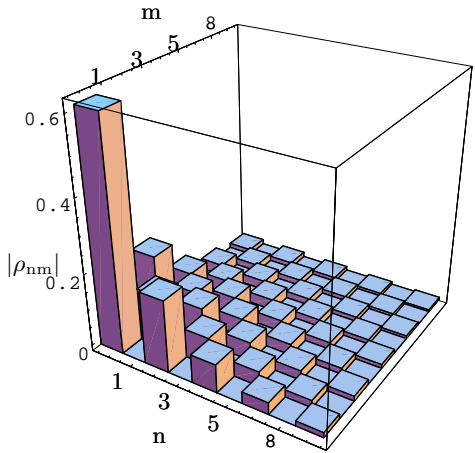}
 \caption[]{\sl\small
   Density matrix elements $\rho_{nm}$ of the stationary QBM-state given by eq. \eqref{rhonm} for mean occupation number $\langle n\rangle =1$ ($\Gamma=100\omega_0$, $\gamma\approx2\omega_0$) at low bath temperature $T=10^{-3}\hbar\omega_0/k$. The stationary QBM-state is a mixed state with purity $\mu<1$.}
 \end{center}
\end{figure}
This is the case when subsystem and environment are entangled and
therefore the ground state does not factorize into a product of a
system wave function and a bath wave function. Formally it is a
consequence of the fact that the Hamiltonian $H_s$ does not commute
with the Hamiltonian $H$ of the total system. The mean energy of an
oscillator coupled to a heat bath which is given by the expression
\begin{eqnarray}\label{Hmittel}
\langle H_s\rangle&=&\mbox{Tr}[\rho_sH_s]=\frac{1}{2m}\langle
p^2\rangle+\frac{1}{2}m\omega_0^2\langle q^2\rangle\\
&=&\frac{m\hbar}{2\pi}\int_0^{\infty}d\omega\,(\omega_0^2+\omega^2)\coth\left(\frac{1}{2}\beta\hbar\omega\right)\mbox{Im}\{\tilde\chi(\omega)\}\nonumber
\end{eqnarray}
is greater than its ground state energy for $T\to 0$:
\begin{equation}\label{Hmittel0}
\langle H_s\rangle_{T=0}=\frac{m\hbar}{2\pi}\int_0^{\infty}d\omega\,(\omega_0^2+\omega^2)\mbox{Im}\{\chi(\omega)\}>\frac{1}{2}\hbar\omega_0.
\end{equation}
The probability to measure the harmonic oscillator in the ground
state or an excited state may be found by considering the diagonal
matrix elements $\rho_{nn}$ of the density matrix $\rho_s$ in the
energy basis
\begin{equation}\label{rhonm}
\rho_{nm}=\langle n|\rho_s|m\rangle=\int dqdq'\langle
n|q\rangle\langle q|\rho_s|q'\rangle\langle q'|m\rangle,
\end{equation}
where $\langle q|n\rangle=\left(\sqrt{\pi}2^nn!\right)^{-1/2}\sqrt{b}H_n(bq)e^{-b^2q^2/2}$
are the wave functions of the harmonic oscillator with Hermite
polynomials $H_n$ and $b=\sqrt{m\omega_0/\hbar}$. The diagonal elements can then be expressed explicitly by
\begin{equation}\label{rhonnQBM}
\rho_{nn}=\sqrt{\frac{4}{D}}(d^2-a^2)^{n/2}P_n\left[\frac{d}{\sqrt{d^2-a^2}}\right],
\end{equation}
where $P_n$ are the Legendre polynomials of order $n$ and the dimensionless variables $x=2b^2\langle q^2\rangle$, $y=2\langle p^2\rangle/(\hbar^2b^2)$, $D=1+x+y+xy$, $a=(y-x)/D$ and $d=(xy-1)/D$ have been introduced. \cite{buettiker1}.
\begin{figure}[t]
\begin{center}
\psfrag{n}[c]{n}\psfrag{m}[c]{m} \psfrag{r}[c]{$|\rho_{\rm nm}|$}
\psfrag{2}[c]{\rm 1}\psfrag{4}[c]{3}\psfrag{6}[c]{\rm 5}
 \includegraphics[width=6cm,clip]{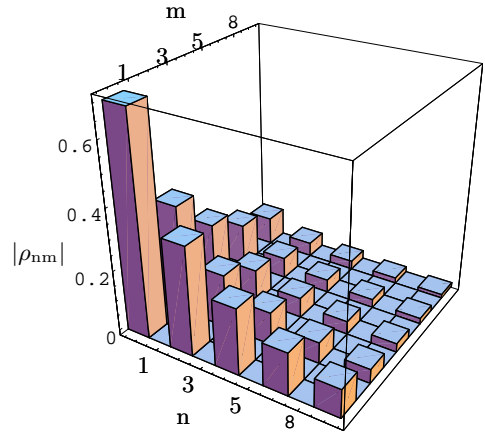}
 \caption[]{\sl\small
   Density matrix elements $\rho_{nm}$ of a squeezed vacuum state $|\xi\rangle=e^{\xi^*a^2/2-\xi(a^{\dagger})^2/2}|0\rangle$ with mean occupation number $\langle n\rangle=\sinh^2|\xi|=1$. The squeezed vacuum state is a pure state with purity $\mu=1$. }
 \end{center}
\end{figure}
As can be seen from figure 1 the density matrix $\rho_{nm}$ is not
strictly diagonal in this basis because the wave functions of the
unperturbed oscillator are not the eigenfunctions of $\rho_s$.
Furthermore, the eigenvalues $p_n$ of $\rho_s(q,q')$ and its
eigenfunctions
$f_n=\left(\sqrt{\pi}2^nn!\right)^{-1/2}\sqrt{c}H_n(cq)e^{-c^2q^2/2}$
with $c=[\langle p^2 \rangle /(\hbar^2 \langle q^2\rangle)]^{1/4}$
are obtained as solution of the problem $\int
dq'\rho(q,q')f_n(q')=p_n f_n(q)$ \cite{NA1}. Making use of the purity
\begin{equation}\label{purity}\mu=\mbox{Tr }\rho_s^2=\int dqdq'\langle q|\rho_s|q'\rangle\langle q'|\rho_s|q\rangle=
\frac{\hbar/2}{\sqrt{\langle q^2\rangle \langle
p^2\rangle}}\end{equation} the eigenvalues can be expressed by
\begin{equation}\label{pn}
p_n=2\mu/(1+\mu)\left[(1-\mu)/(1+\mu)\right]^n.\end{equation}
Figure 1 visualizes the distribution of the density matrix elements in the energy eigenbasis of the QBM-state at low temperatures. One can see that the density matrix in energy eigenbasis is not strictly diagonal and that the distribution on the diagonal deviates from the Boltzmann distribution. This is due to the interaction which correlates system and environment and additionally leads to a squeezed-like state. This becomes obvious when comparing figure 1 to figure 2 which shows the distribution of a squeezed vacuum state. Unlike the squeezed vacuum state, the stationary QBM-state is not a pure state with minimal uncertainty but a mixed state even at zero bath temperature. Diagonalizing the density matrix \eqref{roheq} does not just yield one eigenvalue unequal to zero ($p_0=1$, $p_n=0$ for $n>0$), but also $p_n>0$ for $n>0$. Thus, for non-weak system bath interaction with $\gamma>0$ and zero temperature the purity is $\mu<1$ and the von Neumann entropy $S_v(\rho_s)>0$. Despite the same occupation number $\langle n\rangle$ in fig. 1 und 2, the purity and the von Neumann entropy are different in both cases.\\
\begin{figure}[t]
\begin{center}
\psfrag{x}[c]{\small$(\Delta \tilde q)^2$}\psfrag{p}[c]{\small$(\Delta \tilde p)^2$}
 \includegraphics[width=7cm,clip]{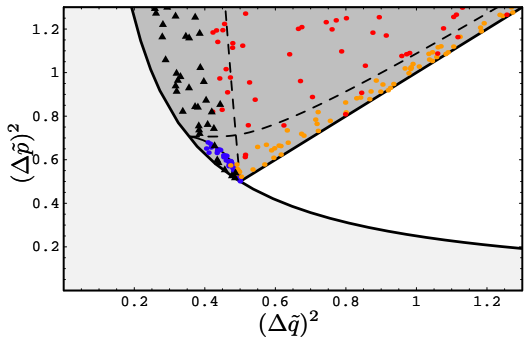}
 \caption[]{\sl\small
   Possible pairs of the normalized QBM-state variances $(\Delta \tilde q)^2=(\Delta q)^2/\sigma_0$  and $(\Delta\tilde p)^2=(\Delta p)^2\frac{4\sigma_0^2}{\hbar^2}$. Physical states lie above the hyperbola which characterizes the set of squeezed vacuum states. Thermal states with Boltzmann distribution are given by bisecting line. The data points are QBM-states for randomly chosen combinations of parameters $(\gamma,\Gamma,T)$. }
 \end{center}
\end{figure}

The environment-induced squeezing increases the uncertainty product $\Delta q\Delta p$ compared to a thermal state. This can be seen from fig. 3 which shows possible and feasible combinations of the second moments of $p$ and $q$. Physical states lie above the hyperbola which characterizes the set of squeezed vacuum states. Thermal states with Boltzmann distribution are given by the bisecting line. The data points in the gray shades area are QBM-states for randomly chosen combinations of parameters $(\gamma,\Gamma,T)$. Possible pairs of the mean occupation number $\bar{n}$ and its variance $(\Delta n)^2$ for arbitrary combinations of $\gamma$, $\Gamma$ and $T$ in the QBM model are bounded from below by the variance of a thermal state $\langle n\rangle(\langle n\rangle+1)$ and from above by the variance of a squeezed vacuum state $2\langle n\rangle(\langle n\rangle+1)$. This is illustrated by figure 4.\\

The statistical properties of the stationary QBM-state hence are a consequence of the thermal noise, environment-induced squeezing and existing quantum correlations between system and bath. The distribution of the diagonal elements in the energy eigenbasis thus deviates from the Boltzmann distribution. Figure 5 visualizes the difference between eq. \eqref{pn} and $\rho_{nn}$ given by eq. \eqref{rhonm}. The population of higher levels is increased while the population of lower levels is reduced compared to a thermal distribution.
\begin{figure}[t]
\psfrag{x}[c]{\small$\langle n\rangle$}\psfrag{y}[c]{\small$(\Delta n)^2$}
\psfrag{h1}[c]{\small $2\langle n\rangle(\langle n\rangle+1)$}\psfrag{h2}[c]{\small$\langle n\rangle(\langle n\rangle+1)$}
\begin{center}
\psfrag{n}[c]{n}\psfrag{m}[c]{m} \psfrag{r}[c]{$\rho_{\rm nm}$}
\psfrag{2}[c]{\rm 1}\psfrag{4}[c]{3}\psfrag{6}[c]{\rm 5}
 \includegraphics[width=7cm,clip]{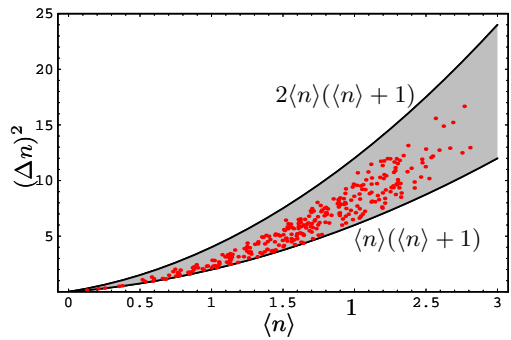}
 \caption[]{\sl\small
   Possible pairs of the QBM-state mean occupation number $\bar{n}$ and its variance $(\Delta n)^2$. The gray shaded area is bounded from below by the variance of a thermal state $\langle n\rangle(\langle n\rangle+1)$ and from above by the variance of a squeezed vacuum state $2\langle n\rangle(\langle n\rangle+1)$. }
 \end{center}
\end{figure}

\subsection{Reduced density matrix and von Neumann entropy}
Since the probability to find the system in an excited state does
not vanish even at zero temperature, the density operator $\rho_s$
does not reduce to the projection onto the non-degenerate ground
state of the system $H_s$ and thus does not describe a pure state
with statistical entropy equal to zero. The uncertainty about the
measurement outcome is expressed by the von Neumann entropy
$S_{v}(\rho_s)$ of the reduced density matrix $\rho_s$ which is in
the case of a Gaussian state given by \cite{agarwal, NA1, serafini}:
\begin{eqnarray}\label{SvN}
S_{v}(\rho_s)&=&-k \mbox{Tr}[\rho_s\ln \rho_s]=-k\sum\limits_np_n\ln
p_n\\&=&k\frac{1-\mu}{2\mu}\ln\frac{1+\mu}{1-\mu}-k\ln
\frac{2\mu}{1+\mu}\nonumber.
\end{eqnarray}
Introducing the effective frequency and mass \cite{weiss}
\begin{equation}
\omega_{\rm
eff}=\frac{2}{\hbar\beta}\mbox{arcoth}\left(\mu\right),\quad
m_{\rm eff}=\frac{\sqrt{\langle p^2\rangle/\langle
q^2\rangle}}{\omega_{\rm eff}},
\end{equation}
the density matrix can be written in the form
\begin{equation}
\rho_s=Z_{\rm eff}^{-1}\exp\left(-\beta H_{\rm eff}\right),
\end{equation}
with effective partition function $Z_{\rm
eff}^{-1}=2\sinh\left(\omega_{\rm eff}\hbar\beta/2\right)$ and
effective Hamiltonian $H_{\rm eff}=\frac{p^2}{2m_{\rm
eff}}+\frac{1}{2}m_{\rm eff}\omega_{\rm eff}^2q^2$. The von Neumann
entropy and its eigenvalues $p_n$ can than be expressed by
\begin{equation}
S_v(\rho_s)=\frac{\beta\hbar\omega_{\rm
eff}}{e^{\beta\hbar\omega_{\rm
eff}}-1}-\ln\left(1-e^{-\beta\hbar\omega_{\rm eff}}\right),\end{equation}
and $
p_n=Z_{\rm eff}^{-1}e^{-\beta E_n}$ with $E_n=\hbar\omega_{\rm
eff}\left(n+1/2\right)$. However, notice that $S_v(\rho_s)\neq S_{\rm eff}=\frac{\partial }{\partial T}(kT\ln Z_{\rm eff})$ due to the temperature dependence of $\omega_{\rm eff}$. Since $\omega_{\rm eff}\to 0$ for $T\to 0$ the effective energy
levels $E_n$ become very narrowly spaced near zero temperature
indicating that the ground state is a mixture. $S_v(\rho_s)$ is the minimal amount of classical information that is
accessible through measurements on the quantum state $\rho_s$
\cite{nielsen} and is thus smaller than the Shannon entropy $\sum_n
\rho_{nn}\log\rho_{nn}$ with matrix elements $\rho_{nn}$ given in
the basis of the unperturbed eigenfunctions of the harmonic
oscillator (which are not the eigenfunctions of $\rho_s$).
\begin{figure}[t]
\psfrag{t}[c]{n} \psfrag{x}[c]{$\rho_{nn}$ vs.
$p_n$}
\begin{center}
\psfrag{n}[c]{n}\psfrag{m}[c]{m} \psfrag{r}[c]{$\rho_{nn}$ vs.
$p_n$}  \includegraphics[width=7.5cm,clip]{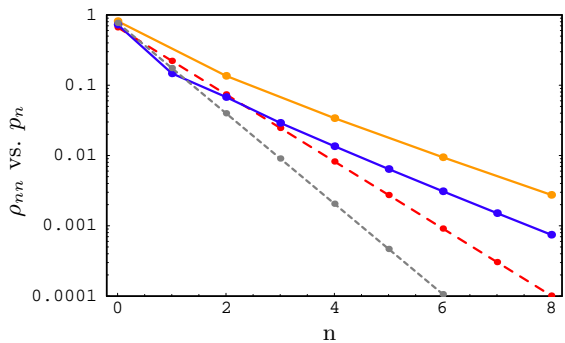}
 \caption[]{\sl\small
   Distribution of the diagonal density matrix elements $\rho_{nn}$ of a QBM-state with $\langle n\rangle =1/2$ (blue line) compared to a squeezed vacuum state (yellow line) and a thermal state (red line) with the same occupation number (on a logarithmic scale). The distribution of the eigenvalues $p_n$ given by eq. \eqref{pn} is the dashed gray line. ($\Gamma=100\omega_0$, $\gamma=0.93\omega_0$).}
 \end{center}
\end{figure}
\section{Comparison of entropies}
The von Neumann entropy $S_v$ of the subsystem $\rho_s$ and the
thermodynamic entropy $S(T)$ of the Brownian particle deviate from
each other as one can see from fig. 6:
\begin{equation}
S_{v}(\rho_s) \geq S(T) \quad\mbox{for}\quad \gamma >0
\end{equation}
Even at $T\to 0$ the von Neumann entropy $S_{v}$ is larger than zero
while its thermodynamic entropy vanishes approximately linearly with
$T$. This effect of $S_{v,T\to0}\neq0$ which has been extensively
discussed in ref. \cite{NA1, NA2} is due to the entanglement between
subsystem and bath which prevents the subsystem from reaching a pure
state for $T=0$.  It is just in the weak coupling limit $\gamma \to
0$ that the von Neumann entropy \eqref{SvN} and the thermodynamic
entropy \eqref{Sthermo} become equal and tend towards the entropy
$s(\omega_0,T)$ of a harmonic oscillator in a heat bath at
temperature $T$, characterized by the Gibbs distribution.\\
The reason for the deviation in entropies may be seen in the fact
that the density matrix $\rho_s$ cannot be written in canonical form
\begin{equation}
\rho_s \neq \frac{1}{\mathcal{Z}}\exp\left(-\beta H_s\right).
\end{equation}
The statistical entropy can be identified with the thermodynamic
entropy just if $\rho=Z^{-1}\exp(-\beta H_s)$ is the canonical
density matrix with $Z=\mbox{Tr}[\exp(-\beta H_s)]$, which means
\begin{eqnarray}
S_{v}(\rho)&=&-\mbox{Tr}[\rho\ln\rho]=\beta\mbox{Tr}[\rho
H_s]+\mbox{Tr}[\rho\ln Z]\nonumber\\
&=&\beta\langle H_s\rangle+\ln Z=\beta(U-F)=S(T).
\end{eqnarray}
In the same manner, the heat $\delta Q$ connected to the
thermodynamic entropy by the Clausius equality $\delta Q=TdS$ (for
quasi-static processes) can only be associated with the part of $dU$
representing a statistical redistribution in phase space
$\mbox{Tr}[d\rho_s H_s]$ if $\rho_s$ is given in canonical form:
\begin{eqnarray}
dS_v&=&-\mbox{Tr}[d\rho\ln\rho]=\mbox{Tr}[d\rho\ln
Z]+\beta\mbox{Tr}[d\rho H_s]\nonumber\\
&=&\beta\mbox{Tr}[d\rho H_s]=\beta\delta Q=dS(T).
\end{eqnarray}
While the choice of the entropy \eqref{Sthermo} implies that this
thermodynamic entropy of the Brownian oscillator $S(T)$ and the
(unperturbed) bath $S_v(\rho_b)$ are additive -- since all
correlations between bath and oscillator are included in the
subsystems entropy -- the von Neumann entropy \eqref{SvN} for the
system and the bath entropy are non-additive
\begin{equation}
S_{v}(\rho_s)+ S_{v}(\rho_b)-I_{sb}= S_{v}(\rho_{\rm tot})=
S(T)+S_{v}(\rho_b),
\end{equation}
with non-zero mutual information/entropy $I_{sb}$ measuring the
correlations between the parts $S$ and $B$ of the entangled system
$\rho_{\rm tot}\neq\rho_s\otimes\rho_b$. The conditional entropy
$S(S|B)=S_{v}(\rho_s)-I_{sb}$ can be identified with the
thermodynamic entropy \eqref{Sthermo}. This entropy measures the
uncertainty about the state of $S$ if the state of $B$ is known
\cite{nielsen}. It is the shift in the total von Neumann entropy
$S_{v}(\rho_{\rm tot})-S_v(\rho_b)$ due to the presence of the
Brownian particle and thus differs strongly from the von Neumann
entropy $S_v(\rho_s)$ of the particle alone.
\begin{figure}[t]
\begin{center}
\psfrag{x}[c]{$kT/\hbar\omega_0$}\psfrag{y}[c]{$S_{v}(\rho_s)/k$, $S(T)/k$}
 \includegraphics[width=7.5cm,clip]{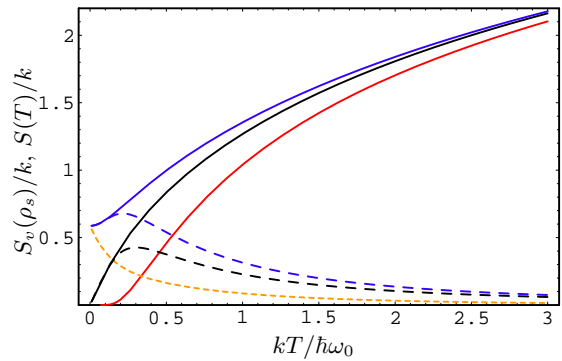}
 \caption[]{\sl\small
   Temperature dependence of the von Neumann entropy $S_{v}(\rho_s)$ (blue line), thermodynamic entropy $S(T)$ (dark line)
   and function $s(\omega,T)$ (red line) for system-bath-coupling $\gamma=10m\omega_0^2/\Gamma=1$ compared to the entropy
   of the uncoupled harmonic oscillator $s(\omega,T)$ (dashed line). Parameters are $\Gamma=10\omega_0$ and $\gamma\approx2.43\omega_0$, thus having $\langle H_s\rangle_{T=0}=\hbar\omega_0$.}
 \end{center}
\end{figure}
\section{Cyclic thermodynamic processes}
The theoretical result of a mean energy $\langle H_s\rangle_{T\to 0}>\frac{1}{2}\hbar\omega_0$ has raised the question whether this is a contradiction to the second law: if the particle is subtracted from the bath the difference in energy could be used to perform work. A cyclic process could be constructed that receives energy for work generation from a single heat bath. That such a construction of a perpetuum mobile of the second kind is not possible was recently shown by Ford and O'Connell \cite{ford0a} for the QBM-model with Lorentzian bath spectral density and by Kim and Mahler \cite{mahler} for arbitrary bath spectra. The basic argument is the principle of minimal work which states that the minimal amount of work that is needed to bring the system from one thermodynamic state to another of the same temperature, is equal to the difference in the free energy of these two thermodynamic states \cite{landau}. If the Brownian oscillator is added to or subtracted from the bath the difference in free energy is equal to $F(T)$ in eq. \eqref{Fthermo}. This can be seen from the following considerations. The free energy of the total system is given by the sum
\begin{equation}\label{ftot}
F_{\rm tot}(T,\gamma)=\sum_k\,f(\omega_k',T),
\end{equation}where the eigenfrequencies $\omega_k'$ of the coupled system follow as the roots of a transcendental eigenvalue equation \cite{NA1} and $f(\omega_k',T)=kT\ln 2\sinh(\frac{1}{2}\beta\hbar\omega_k')$. For $\omega_k\gg\Delta$ the initial frequencies $\omega_k=k\Delta$ get shifted to
\begin{equation}
\omega_k'=\omega_k-\frac{\Delta}{\pi}\phi(\omega_k)\quad\mbox{for}\quad\omega_k\gg\Delta,
\end{equation}
where $\phi(\omega)$ is related to the susceptibility $\tilde\chi(\omega)$ by
\begin{eqnarray}
\phi(\omega)&=&\int_0^{\omega}d\nu\,\mbox{Im}\left\{\frac{d\ln\tilde
\chi(\nu)}{d\nu}\right\}\nonumber\\
&=&\sum\limits_{j=1}^3\arctan\frac{\omega}{\lambda_j}-\arctan\frac{\omega}{\Gamma}.
\end{eqnarray}
Using the identity (for small $\Delta$)
\begin{equation}
\sum_kf(\omega_k')\rightarrow\frac{1}{\Delta}\int_0^{\infty}d\omega_k'\,\frac{d\omega_k}{d\omega_k'}\,f(\omega_k'),
\end{equation}
the total free energy \eqref{ftot} can be written as
\begin{eqnarray}
F_{\rm tot}(T,\gamma)&=&\int_0^{\infty}d\omega\left[\frac{1}{\Delta}+\frac{1}{\pi}\frac{d\phi(\omega)}{d\omega}\right]\,f(\omega,T)\nonumber\\
&=&\sum_k f(\omega_k,T)+\frac{1}{\pi}\int_0^{\infty}d\omega f(\omega,T)\frac{d\phi(\omega)}{d\omega}\nonumber\\
&=&F_b(T,\gamma=0)+F(T),
\end{eqnarray}
where $F_b(T,\gamma=0)$ is the free energy of the undisturbed bath in absence of the particle. Thus, the shift in free energy due to the coupling of the central oscillator to the bath is $F(T)$. The related increase in internal energy is then expressed by $U(T)=F(T)+TS(T)$. However, the mean energy of the oscillator is given by \eqref{Hmittel}. To ensure the validity of the second law at zero temperature, the work which has to be performed has to be larger than the mean energy. Because of
$F(0)-\langle H\rangle_{T=0}\geq0$
this is the case \cite{ford0a, mahler}. Since at $T\to 0$ all the employed work is converted into internal energy $U(0)$ of the total system, the difference can be interpreted as interaction energy $U_{\rm int}$ (which is different from $\langle H_{\rm int}\rangle$!). Due to the infinite number of bath modes this increase in internal energy does not raise the bath temperature. The interaction energy is given by the partial differentiation of the free energy \eqref{Fthermo} with respect to the cut-off frequency \cite{NA1}
\begin{equation}U_{\rm int}=U-\langle H_s\rangle=\Gamma\frac{\partial F}{\partial
\Gamma}.\end{equation} With these considerations the zero temperature behavior of the thermodynamic entropy becomes plausible.  The free energy $F(0)$ is completely converted into internal energy $U(0)$, that is $\langle H_s\rangle$ and $U_{\rm int}$. This increase in internal energy is not accompanied by a flow of heat or a change in entropy at zero temperature, thus $S(0)=0$ and $S_v(\rho_{\rm tot})=S_v(\rho_b(0))$. On the other hand, variations of the oscillator parameters $d\omega_0$, $dm$ are related to a change in entropy $dS(T)$ and a heat flow respectively that can be
calculated by comparing the total differential
\begin{eqnarray}
dU&=&d\langle H_s\rangle +dU_{\rm int}\nonumber\\
&=&{\rm Tr}[\rho_sdH_s]+{\rm Tr}[H_sd\rho_s]+dU_{\rm int}\nonumber\\
&=&\delta W_s+\delta Q_s+dU_{\rm int}
\end{eqnarray}
with the total differential $dU=dF+TdS$. A short calculation
reveals that $dF=\delta W_s:={\rm
Tr}[\rho_sdH_s]$  and thus one has
\begin{equation}
\delta Q:=TdS(T)=\delta Q_s+dU_{\rm int}.
\end{equation}
Defining the total heat flow $\delta Q$ accompanying a variation of
the entropy $S(T)$ by the Clausius (in)equality $\delta Q=TdS(T)$ it
becomes clear that a change in the thermodynamic entropy of the
Brownian oscillator does not just lead to a heat flow expressed by a
redistribution in phase space of the subsystem ${\rm
Tr}[H_sd\rho_s]=:\delta Q_s$ but also to a flow of heat to the cloud
of surrounding bath modes which is $\delta Q_s=-dU_{\rm int}$ for $T\to 0$.
For a recent discussion on problems arising from a definition of local heat and work
in bipartite quantum systems see e.g. ref. \cite{mahler2}.

\section{Consequences for quantum information theory}
From an information-theoretical point of view, these considerations
become important if statements about the information capacity of a
quantum system are connected to thermodynamic properties. We may
cite the Landauer principle \cite{landauer, bennett} as an example
here. This principle is based on the Clausius inequality and states
that many-to-one-operations like erasure of information requires the
dissipation of energy. Deleting one bit of information of the
systems memory is accompanied by a released amount of heat of at
least $kT\ln 2$:
\begin{equation}\label{LP}
\left|\frac{\delta Q}{dS/(\ln 2)}\right|\geq kT\ln 2.
\end{equation}
In our case we apply the Landauer principle to quantum
systems in which classical information is encoded. Since the the
optimal erasure procedure, i.e. the one that creates the least
amount of heat is the one where the quantum measurements are made in
the basis of the eigenstates of $\rho_s$, the relevant entropy is
the von Neumann entropy $S_v(\rho_s)$ of the system. The erasure of
information is connected to a reduction of entropy, and thus cannot
be realized in a closed system. Therefore the information-carrying
system has to be coupled to the environment. In a low temperature
environment the coupling may be relatively strong compared to the
thermal energy and system and bath become correlated.\\

Of course, the customary understanding of Landauer's principle is
that it applies to situations in which the information-bearing
system is uncorrelated with the environment both initially and
finally. Classically Landauer's principle holds as long as the
second law of thermodynamics is applicable in the given situation.
With existing correlations between system and environment both
principles - the Landauer bound as well as the second law - are not
applicable because the basic requirement of additivity is not
fulfilled. So, why considering the Landauer principle in this
context? The point is, that in our case of quantum Brownian motion
the second law of thermodynamics holds as has been pointed out in
the literature \cite{ford0a, ford0b, haenggi2} while the Landauer
principle may not be applicable as long as the von Neumann entropy
of the reduced density matrix is seen as the relevant quantity to
describe the classical information content of the quantum state of
the damped harmonic oscillator. The reason for this is, that the von
Neumann entropy and the thermodynamic entropy may deviate from each
other for mesoscopic quantum systems that are sufficiently strong
entangled with their environment. Since the statistical entropy and
the heat are defined separately they do not necessarily have to
fulfill the Clausius inequality \cite{NA2, hoebu}. A second reason
for considering the Landauer principle in this context is, that we
are able to quantify the deviations from the Landauer bound that are
resulting from the quantum correlations between the information
carrying subsystem and its surrounding bath. At least it might be
interesting to see in which way deviations from the Landauer
principle occur if the assumption of weak interaction between system
and bath is gradually abandoned.\\

As an example for our system we show in figure 7 the temperature
dependence of the heat exchange $\delta Q$ divided by the entropy
change, in this case of the von Neumann entropy $dS_v$ (in bits).
The heat is defined by $\delta Q=TdS(T)$ for quasi-static variations
of the oscillator frequency $d\omega_0$. Since $\delta Q_{s}$ and
$\delta Q_{\rm int}$ can have different signs, the heat flow $\delta
Q$ is lowered compared to the term $\delta Q_s$. Thus, as it can
clearly be seen, this ratio for small temperatures is {\sl below}
the Landauer bound of $kT\ln 2$. Actually at low temperature and
high cut-off frequency this quantity starts out quadratically as a
function of temperature instead of linearly, which is due to the
fact that the von Neumann entropy term is larger than its canonical
form (thermodynamic entropy) and therefore the Landauer principle in
this case gives only an upper bound to the erasure of information.
\begin{figure}[t]
\begin{center}
\psfrag{T}[c]{$kT/\hbar\omega_0$}\psfrag{dS}[c]{ratio $|\delta
Q/dS_{v}\ln2|$ }\psfrag{1}[c][r][1][30]{$kT\ln 2$}
 \includegraphics[width=7.5cm,clip]{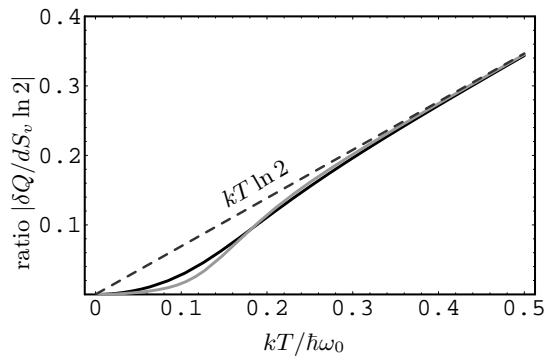}
 \caption[]{\sl\small
   Temperature dependence of the ratio $|\delta Q/dS_{v}\ln2|$ (in bits) with the heat defined by $\delta Q=TdS(T)$ for quasi-static variations of the oscillator frequency
   $d\omega_0$. The system-bath-couplings are chosen to be $\gamma=0.1\omega_0$ (dark line) and $\gamma=0.5\omega_0$ (gray line). At low T deviations from the Landauer bound $kT\ln 2$ (dashed line) occur. Here $\Gamma=10\omega_0$.}
 \end{center}
\end{figure}

\section{summary and conclusions}
In this paper we have studied the statistical and thermodynamic properties of the stationary QBM-state.
We have shown that an identification of the thermodynamic entropy of
the damped quantum oscillator with its von Neumann entropy is not
valid except in the weak coupling limit. We have chosen Landauer's
principle to determine these deviations
qualitatively and quantitatively.\\
The statistical entropy associated with the stationary quantum state
is the von Neumann entropy. The main features of the subsystem (e.g.
mixed state at zero temperature, accessible information by
measurements) are captured by the von Neumann entropy of the reduced
density matrix. As a well established measure of entanglement and
decoherence in open quantum systems it is commonly in use -- but
interpretational problems may arise if it is related to
thermodynamic quantities. \\The thermodynamic entropy is the correct
choice when cyclic processes are considered (e.g. to show that no
work can be extracted from a single thermal bath \cite{ford0a, pusz,
lenard, NA4}). The exchange of heat between the system and the
environment during quasi-static processes is related to the
thermodynamic entropy by the Clausius (in)equality. However, it is
misleading to speak of {\sl the} entropy of the Brownian particle as
we have pointed out. The thermodynamic entropy can just be
identified with the statistical entropy when $\rho_s$ takes the form
of the canonical density matrix. This is just the case for
negligible interaction between subsystem and environment. As long as
the application of the von Neumann entropy is restricted to an
information-theoretical context and the thermodynamic entropy is
applied when examining the validity of the thermodynamic laws no
interpretational problems will arise. The situation is different in
cases where statements about the information content of a system are
connected to thermodynamic properties as has been demonstrated by
means of the Landauer principle. In this case the entanglement
between the subsystem and the bath is responsible for the fact that
the erasure of information -- measured by the decrease in the von
Neumann entropy of the reduced density matrix -- is accompanied by a
released amount of heat that may be below the Landauer bound.

{\sl Acknowledgement --} We would like to thank P. H\"anggi for valuable comments on the subject of this paper.


\begin{thebibliography}{35}
\expandafter\ifx\csname natexlab\endcsname\relax\def\natexlab#1{#1}\fi
\expandafter\ifx\csname bibnamefont\endcsname\relax
  \def\bibnamefont#1{#1}\fi
\expandafter\ifx\csname bibfnamefont\endcsname\relax
  \def\bibfnamefont#1{#1}\fi
\expandafter\ifx\csname citenamefont\endcsname\relax
  \def\citenamefont#1{#1}\fi
\expandafter\ifx\csname url\endcsname\relax
  \def\url#1{\texttt{#1}}\fi
\expandafter\ifx\csname urlprefix\endcsname\relax\def\urlprefix{URL }\fi
\providecommand{\bibinfo}[2]{#2}
\providecommand{\eprint}[2][]{\url{#2}}

\bibitem[{\citenamefont{Weiss}(1999)}]{weiss}
\bibinfo{author}{\bibfnamefont{U.}~\bibnamefont{Weiss}},
  \emph{\bibinfo{title}{Quantum Dissipative Systems}}
  (\bibinfo{publisher}{World Scientific}, \bibinfo{address}{Singapore},
  \bibinfo{year}{1999}).

\bibitem[{\citenamefont{Breuer and Petruccione}(2003)}]{breuer}
\bibinfo{author}{\bibfnamefont{H.-P.} \bibnamefont{Breuer}} \bibnamefont{and}
  \bibinfo{author}{\bibfnamefont{F.}~\bibnamefont{Petruccione}},
  \emph{\bibinfo{title}{The Theory of Open Quantum Systems}}
  (\bibinfo{publisher}{Oxford University Press}, \bibinfo{address}{Oxford},
  \bibinfo{year}{2003}).

\bibitem[{\citenamefont{Dittrich et~al.}(1998)\citenamefont{Dittrich, H\"anggi,
  Ingold, Kramer, Sch\"on, and Zwerger}}]{dittrich}
\bibinfo{author}{\bibfnamefont{T.}~\bibnamefont{Dittrich}},
  \bibinfo{author}{\bibfnamefont{P.}~\bibnamefont{H\"anggi}},
  \bibinfo{author}{\bibfnamefont{G.-L.} \bibnamefont{Ingold}},
  \bibinfo{author}{\bibfnamefont{B.}~\bibnamefont{Kramer}},
  \bibinfo{author}{\bibfnamefont{G.}~\bibnamefont{Sch\"on}}, \bibnamefont{and}
  \bibinfo{author}{\bibfnamefont{W.}~\bibnamefont{Zwerger}},
  \emph{\bibinfo{title}{Quantum Transport and Dissipation}}
  (\bibinfo{publisher}{Wiley-VCH}, \bibinfo{address}{Weinheim},
  \bibinfo{year}{1998}).

\bibitem[{\citenamefont{Caldeira and Leggett}(1983{\natexlab{a}})}]{caldeira1}
\bibinfo{author}{\bibfnamefont{A.~O.} \bibnamefont{Caldeira}} \bibnamefont{and}
  \bibinfo{author}{\bibfnamefont{A.~J.} \bibnamefont{Leggett}},
  \bibinfo{journal}{{\rm Ann. Phys. (N.Y.)}} \textbf{\bibinfo{volume}{{\bf
  149}}}, \bibinfo{pages}{374} (\bibinfo{year}{1983}{\natexlab{a}}).

\bibitem[{\citenamefont{Ford et~al.}(1988{\natexlab{a}})\citenamefont{Ford,
  Lewis, and O'Connell}}]{ford1}
\bibinfo{author}{\bibfnamefont{G.~W.} \bibnamefont{Ford}},
  \bibinfo{author}{\bibfnamefont{J.~T.} \bibnamefont{Lewis}}, \bibnamefont{and}
  \bibinfo{author}{\bibfnamefont{R.~F.} \bibnamefont{O'Connell}},
  \bibinfo{journal}{{\rm Phys. Rev. A}} \textbf{\bibinfo{volume}{{\bf 37}}},
  \bibinfo{pages}{4419} (\bibinfo{year}{1988}{\natexlab{a}}).

\bibitem[{\citenamefont{Nielsen and Chuang}(2000)}]{nielsen}
\bibinfo{author}{\bibfnamefont{M.~A.} \bibnamefont{Nielsen}} \bibnamefont{and}
  \bibinfo{author}{\bibfnamefont{I.~L.} \bibnamefont{Chuang}},
  \emph{\bibinfo{title}{Quantum Computation and Quantum Information}}
  (\bibinfo{publisher}{Cambridge Univ. Press}, \bibinfo{address}{Cambridge},
  \bibinfo{year}{2000}).

\bibitem[{\citenamefont{Braunstein and van Look}(2005)}]{braunstein}
\bibinfo{author}{\bibfnamefont{S.}~\bibnamefont{Braunstein}} \bibnamefont{and}
  \bibinfo{author}{\bibfnamefont{P.}~\bibnamefont{van Look}},
  \bibinfo{journal}{{\rm Rev. Mod. Phys.}} \textbf{\bibinfo{volume}{{\bf 77}}},
  \bibinfo{pages}{513} (\bibinfo{year}{2005}).

\bibitem[{\citenamefont{Cerf et~al.}(2007)\citenamefont{Cerf, Leuchs, and
  Polzik}}]{cerf}
\bibinfo{author}{\bibfnamefont{N.~J.} \bibnamefont{Cerf}},
  \bibinfo{author}{\bibfnamefont{G.}~\bibnamefont{Leuchs}}, \bibnamefont{and}
  \bibinfo{author}{\bibfnamefont{E.~S.} \bibnamefont{Polzik}},
  \emph{\bibinfo{title}{Quantum Information with Continuous Variables of Atoms
  and Light}} (\bibinfo{publisher}{Imperial College Press},
  \bibinfo{address}{London}, \bibinfo{year}{2007}).

\bibitem[{\citenamefont{Sheehan}(2002)}]{sheehan}
\bibinfo{author}{\bibfnamefont{D.~P.} \bibnamefont{Sheehan}},
  \emph{\bibinfo{title}{Quantum Limits to the Second Law}}
  (\bibinfo{publisher}{American Inst. of Physics}, \bibinfo{address}{Melville,
  NY}, \bibinfo{year}{2002}).

\bibitem[{\citenamefont{Nieuwenhuizen et~al.}(2004)\citenamefont{Nieuwenhuizen,
  Spicka, and Keefe}}]{nieuw}
\bibinfo{author}{\bibfnamefont{T.~M.} \bibnamefont{Nieuwenhuizen}},
  \bibinfo{author}{\bibfnamefont{V.}~\bibnamefont{Spicka}}, \bibnamefont{and}
  \bibinfo{author}{\bibfnamefont{P.~D.} \bibnamefont{Keefe}},
  \emph{\bibinfo{title}{Frontiers of Quantum and Mesocopic Thermodynamics}}
  (\bibinfo{publisher}{Elsevier}, \bibinfo{address}{Amsterdam},
  \bibinfo{year}{2004}).

\bibitem[{\citenamefont{Allahverdyan and Nieuwenhuizen}(2002)}]{NA1}
\bibinfo{author}{\bibfnamefont{A.~E.} \bibnamefont{Allahverdyan}}
  \bibnamefont{and} \bibinfo{author}{\bibfnamefont{T.~M.}
  \bibnamefont{Nieuwenhuizen}}, \bibinfo{journal}{{\rm Phy. Rev. E}}
  \textbf{\bibinfo{volume}{{\bf 66}}}, \bibinfo{pages}{036102}
  (\bibinfo{year}{2002}).

\bibitem[{\citenamefont{Ford and O'Connell}(2006)}]{ford0a}
\bibinfo{author}{\bibfnamefont{G.~W.} \bibnamefont{Ford}} \bibnamefont{and}
  \bibinfo{author}{\bibfnamefont{R.~F.} \bibnamefont{O'Connell}},
  \bibinfo{journal}{{\rm Phys. Rev. Lett.}} \textbf{\bibinfo{volume}{{\bf
  96}}}, \bibinfo{pages}{020402} (\bibinfo{year}{2006}).

\bibitem[{\citenamefont{Ford and O'Connell}(2005)}]{ford0b}
\bibinfo{author}{\bibfnamefont{G.~W.} \bibnamefont{Ford}} \bibnamefont{and}
  \bibinfo{author}{\bibfnamefont{R.~F.} \bibnamefont{O'Connell}},
  \bibinfo{journal}{{\rm Physica E}} \textbf{\bibinfo{volume}{{\bf 29}}},
  \bibinfo{pages}{82} (\bibinfo{year}{2005}).

\bibitem[{\citenamefont{H\"anggi and Ingold}(2006)}]{haenggi2}
\bibinfo{author}{\bibfnamefont{P.}~\bibnamefont{H\"anggi}} \bibnamefont{and}
  \bibinfo{author}{\bibfnamefont{G.~L.} \bibnamefont{Ingold}},
  \bibinfo{journal}{{\rm Acta Phys. Pol. B}} \textbf{\bibinfo{volume}{{\bf
  37}}}, \bibinfo{pages}{1537} (\bibinfo{year}{2006}).

\bibitem[{\citenamefont{Ohya and Petz}(2004)}]{ohya}
\bibinfo{author}{\bibfnamefont{M.}~\bibnamefont{Ohya}} \bibnamefont{and}
  \bibinfo{author}{\bibfnamefont{D.}~\bibnamefont{Petz}},
  \emph{\bibinfo{title}{Quantum Entropy and its Use}}
  (\bibinfo{publisher}{Springer}, \bibinfo{address}{Berlin Heidelberg},
  \bibinfo{year}{2004}).

\bibitem[{\citenamefont{Caldeira and Leggett}(1981)}]{caldeira2}
\bibinfo{author}{\bibfnamefont{A.~O.} \bibnamefont{Caldeira}} \bibnamefont{and}
  \bibinfo{author}{\bibfnamefont{A.~J.} \bibnamefont{Leggett}},
  \bibinfo{journal}{{\rm Phys. Rev. Lett.}} \textbf{\bibinfo{volume}{{\bf
  46}}}, \bibinfo{pages}{211} (\bibinfo{year}{1981}).

\bibitem[{\citenamefont{Caldeira and Leggett}(1983{\natexlab{b}})}]{caldeira3}
\bibinfo{author}{\bibfnamefont{A.~O.} \bibnamefont{Caldeira}} \bibnamefont{and}
  \bibinfo{author}{\bibfnamefont{A.~J.} \bibnamefont{Leggett}},
  \bibinfo{journal}{{\rm Physica A}} \textbf{\bibinfo{volume}{{\bf 121}}},
  \bibinfo{pages}{587} (\bibinfo{year}{1983}{\natexlab{b}}).

\bibitem[{\citenamefont{Ford et~al.}(1965)\citenamefont{Ford, Kac, and
  Mazur}}]{ford4}
\bibinfo{author}{\bibfnamefont{G.~W.} \bibnamefont{Ford}},
  \bibinfo{author}{\bibfnamefont{M.}~\bibnamefont{Kac}}, \bibnamefont{and}
  \bibinfo{author}{\bibfnamefont{P.}~\bibnamefont{Mazur}},
  \bibinfo{journal}{{\rm J. Math. Phys.}} \textbf{\bibinfo{volume}{{\bf 6}}},
  \bibinfo{pages}{504} (\bibinfo{year}{1965}).

\bibitem[{\citenamefont{Ford et~al.}(1988{\natexlab{b}})\citenamefont{Ford,
  Lewis, and O'Connell}}]{ford3}
\bibinfo{author}{\bibfnamefont{G.~W.} \bibnamefont{Ford}},
  \bibinfo{author}{\bibfnamefont{J.~T.} \bibnamefont{Lewis}}, \bibnamefont{and}
  \bibinfo{author}{\bibfnamefont{R.~F.} \bibnamefont{O'Connell}},
  \bibinfo{journal}{{\rm J. Stat. Phys.}} \textbf{\bibinfo{volume}{{\bf 53}}},
  \bibinfo{pages}{439} (\bibinfo{year}{1988}{\natexlab{b}}).

\bibitem[{\citenamefont{Ford and Kac}(1987)}]{ford2}
\bibinfo{author}{\bibfnamefont{G.~W.} \bibnamefont{Ford}} \bibnamefont{and}
  \bibinfo{author}{\bibfnamefont{M.}~\bibnamefont{Kac}}, \bibinfo{journal}{{\rm
  J. Stat. Phys.}} \textbf{\bibinfo{volume}{{\bf 46}}}, \bibinfo{pages}{803}
  (\bibinfo{year}{1987}).

\bibitem[{\citenamefont{Gardiner and Zoller}(2000)}]{gardiner}
\bibinfo{author}{\bibfnamefont{C.~W.} \bibnamefont{Gardiner}} \bibnamefont{and}
  \bibinfo{author}{\bibfnamefont{P.}~\bibnamefont{Zoller}},
  \emph{\bibinfo{title}{Quantum Noise}} (\bibinfo{publisher}{Springer Verlag},
  \bibinfo{address}{Berlin Heidelberg New York}, \bibinfo{year}{2000}).

\bibitem[{\citenamefont{H\"anggi and Ingold}(2005)}]{haenggi1}
\bibinfo{author}{\bibfnamefont{P.}~\bibnamefont{H\"anggi}} \bibnamefont{and}
  \bibinfo{author}{\bibfnamefont{G.-L.} \bibnamefont{Ingold}},
  \bibinfo{journal}{{\rm Chaos}} \textbf{\bibinfo{volume}{{\bf 15}}},
  \bibinfo{pages}{026105} (\bibinfo{year}{2005}).

\bibitem[{\citenamefont{Grabert et~al.}(1988)\citenamefont{Grabert, Schramm,
  and Ingold}}]{grabert}
\bibinfo{author}{\bibfnamefont{H.}~\bibnamefont{Grabert}},
  \bibinfo{author}{\bibfnamefont{P.}~\bibnamefont{Schramm}}, \bibnamefont{and}
  \bibinfo{author}{\bibfnamefont{G.-L.} \bibnamefont{Ingold}},
  \bibinfo{journal}{{\rm Phys. Rep.}} \textbf{\bibinfo{volume}{{\bf 168}}},
  \bibinfo{pages}{115} (\bibinfo{year}{1988}).

\bibitem[{\citenamefont{Jordan and B\"uttiker}(2004)}]{buettiker1}
\bibinfo{author}{\bibfnamefont{A.~N.} \bibnamefont{Jordan}} \bibnamefont{and}
  \bibinfo{author}{\bibfnamefont{M.}~\bibnamefont{B\"uttiker}},
  \bibinfo{journal}{{\rm Phys. Rev. Lett.}} \textbf{\bibinfo{volume}{{\bf
  92}}}, \bibinfo{pages}{247901} (\bibinfo{year}{2004}).

\bibitem[{\citenamefont{B\"uttiker and Jordan}(2005)}]{buettiker2}
\bibinfo{author}{\bibfnamefont{M.}~\bibnamefont{B\"uttiker}} \bibnamefont{and}
  \bibinfo{author}{\bibfnamefont{A.~N.} \bibnamefont{Jordan}},
  \bibinfo{journal}{{\rm Physica E}} \textbf{\bibinfo{volume}{{\bf 29}}},
  \bibinfo{pages}{272} (\bibinfo{year}{2005}).

\bibitem[{\citenamefont{Agarwal}(1971)}]{agarwal}
\bibinfo{author}{\bibfnamefont{G.~S.} \bibnamefont{Agarwal}},
  \bibinfo{journal}{{\rm Phys. Rev. A}} \textbf{\bibinfo{volume}{{\bf 3}}},
  \bibinfo{pages}{828} (\bibinfo{year}{1971}).

\bibitem[{\citenamefont{Serafini et~al.}(2005)\citenamefont{Serafini, Paris,
  Illuminati, and De~Siena}}]{serafini}
\bibinfo{author}{\bibfnamefont{A.}~\bibnamefont{Serafini}},
  \bibinfo{author}{\bibfnamefont{M.~G.~A.} \bibnamefont{Paris}},
  \bibinfo{author}{\bibfnamefont{F.}~\bibnamefont{Illuminati}},
  \bibnamefont{and} \bibinfo{author}{\bibfnamefont{S.}~\bibnamefont{De~Siena}},
  \bibinfo{journal}{{\rm J. Opt. B; Quantum Semiclass. Opt.}}
  \textbf{\bibinfo{volume}{{\bf 7}}}, \bibinfo{pages}{R19}
  (\bibinfo{year}{2005}).

\bibitem[{\citenamefont{Kim and Mahler}(2007)}]{mahler}
\bibinfo{author}{\bibfnamefont{I.}~\bibnamefont{Kim}} \bibnamefont{and}
  \bibinfo{author}{\bibfnamefont{G.}~\bibnamefont{Mahler}},
  \bibinfo{journal}{{\rm Eur. Phys. J. B}} \textbf{\bibinfo{volume}{{\bf }54}},
  \bibinfo{pages}{405} (\bibinfo{year}{2007}).

\bibitem[{\citenamefont{Landau and Lifshitz}(1980)}]{landau}
\bibinfo{author}{\bibfnamefont{D.~P.} \bibnamefont{Landau}} \bibnamefont{and}
  \bibinfo{author}{\bibfnamefont{E.~M.} \bibnamefont{Lifshitz}},
  \emph{\bibinfo{title}{Statistical Physics, Part 1}}
  (\bibinfo{publisher}{Pergamon}, \bibinfo{address}{London},
  \bibinfo{year}{1980}).

\bibitem[{\citenamefont{Weimer et~al.}(2005)\citenamefont{Weimer, Henrich,
  Rempp, Schr\"oder and Mahler}}]{mahler2}
\bibinfo{author}{\bibfnamefont{H.}~\bibnamefont{Weimer}},
  \bibinfo{author}{\bibfnamefont{M.~J.} \bibnamefont{Henrich}},
  \bibinfo{author}{\bibfnamefont{F.}~\bibnamefont{Rempp}},
  \bibinfo{author}{\bibfnamefont{H.}~\bibnamefont{Schr\"oder}},
  \bibnamefont{and} \bibinfo{author}{\bibfnamefont{G.}~\bibnamefont{Mahler}},
  \bibinfo{journal}{arXiv:0708.2354}.

\bibitem[{\citenamefont{Landauer}(1961)}]{landauer}
\bibinfo{author}{\bibfnamefont{R.}~\bibnamefont{Landauer}},
  \bibinfo{journal}{{\rm IBM J. Res. Dev.}} \textbf{\bibinfo{volume}{{\bf 5}}},
  \bibinfo{pages}{183} (\bibinfo{year}{1961}).

\bibitem[{\citenamefont{Bennett}(1982)}]{bennett}
\bibinfo{author}{\bibfnamefont{C.~H.} \bibnamefont{Bennett}},
  \bibinfo{journal}{{\rm Int. J. Theor. Phys.}} \textbf{\bibinfo{volume}{{\bf
  21}}}, \bibinfo{pages}{905} (\bibinfo{year}{1982}).

\bibitem[{\citenamefont{Allahverdyan and Nieuwenhuizen}(2001)}]{NA2}
\bibinfo{author}{\bibfnamefont{A.~E.} \bibnamefont{Allahverdyan}}
  \bibnamefont{and} \bibinfo{author}{\bibfnamefont{T.~M.}
  \bibnamefont{Nieuwenhuizen}}, \bibinfo{journal}{{\rm Phys. Rev. E}}
  \textbf{\bibinfo{volume}{{\bf 64}}}, \bibinfo{pages}{056117}
  (\bibinfo{year}{2001}).

\bibitem[{\citenamefont{H\"orhammer and B\"uttner}(2005)}]{hoebu}
\bibinfo{author}{\bibfnamefont{C.}~\bibnamefont{H\"orhammer}} \bibnamefont{and}
  \bibinfo{author}{\bibfnamefont{H.}~\bibnamefont{B\"uttner}},
  \bibinfo{journal}{{\rm J. Phys. A: Math. Gen.}} \textbf{\bibinfo{volume}{{\bf
  38}}}, \bibinfo{pages}{7325} (\bibinfo{year}{2005}), \bibinfo{journal}{arXiv:cond-mat/0408536v2}.

\bibitem[{\citenamefont{Pusz and Woronowicz}(1978)}]{pusz}
\bibinfo{author}{\bibfnamefont{W.}~\bibnamefont{Pusz}} \bibnamefont{and}
  \bibinfo{author}{\bibfnamefont{L.}~\bibnamefont{Woronowicz}},
  \bibinfo{journal}{{\rm Comm. Math. Phys.}} \textbf{\bibinfo{volume}{{\bf
  58}}}, \bibinfo{pages}{273} (\bibinfo{year}{1978}).

\bibitem[{\citenamefont{Lenard}(1978)}]{lenard}
\bibinfo{author}{\bibfnamefont{A.}~\bibnamefont{Lenard}},
  \bibinfo{journal}{{\rm J. Stat. Phys.}} \textbf{\bibinfo{volume}{{\bf 19}}},
  \bibinfo{pages}{575} (\bibinfo{year}{1978}).

\bibitem[{\citenamefont{Allahverdyan and Nieuwenhuizen}(2000)}]{NA4}
\bibinfo{author}{\bibfnamefont{A.~E.} \bibnamefont{Allahverdyan}}
  \bibnamefont{and} \bibinfo{author}{\bibfnamefont{T.~M.}
  \bibnamefont{Nieuwenhuizen}}, \bibinfo{journal}{{\rm Phys. Rev. Lett.}}
  \textbf{\bibinfo{volume}{{\bf 85}}}, \bibinfo{pages}{1799}
  (\bibinfo{year}{2000}).

\end{thebibliography}
\end{document}